\documentclass{article}
\usepackage{graphics}
 \usepackage{graphicx}

 \usepackage{epsfig}

\usepackage{amssymb}
\usepackage{amsmath}
 \usepackage{amsthm}

\begin{document}

\title{Thermoelectric effects in a double quantum dot system weakly coupled to ferromagnetic leads}

\author{M.˜ Bagheri Tagani,
        H.˜ Rahimpour Soleimani,\\
        \small{Department of physics, University of Guilan, P.O.Box 41335-1914, Rasht, Iran}}

\maketitle

\begin{abstract}
Thermoelectric effects through a serial double quantum dot system
weakly coupled to ferromagnetic leads are analyzed. Formal
expressions of electrical conductance, thermal conductance, and
thermal coefficient are obtained by means of Hubbard operators.
Results show that although thermopower is independent of
polarization of leads, figure of merit is reduced by increase of
polarization. Influences of temperature and interdot tunneling on
the figure of merit are also investigated, and observed that
increase of the interdot tunneling strength results in reduction
of the figure of merit. Effect of  temperature on the thermal
conductance is also analyzed.
\end{abstract}

\section{Introduction}  %%%  section command 1.
Thermopower through devices fabricated using nanotechnology has
been increasingly studied in recent
years~\cite{Dubi,Andreev,Vashaee,Finch,Dubi1,Soulier,Nozaki,Kuo,Moca,Singh,Hochbaum}.
Recent advances in the fabrication of nanodevices and the
information showing nanodevices can work as thermal generators
with high efficiency~\cite{Balandin,Khitun,Mazzamuto}, have
increased the importance of such studies. Indeed, violation of the
Wiedemann-Franz low~\cite{Kubala,Garg}, ratio of the electrical
conductance to the thermal conductance is proportional to the
operating temperature, in nanodevices results in increase of the
thermal efficiency and the figure of merit,
$ZT=\frac{S^2G_VT}{\kappa}$, where $S$ is the thermopower, $G_V$
and $T$ denote the electrical conductance and the temperature,
respectively. $\kappa=\kappa_c+\kappa_{ph}$ is composed of the
electrical and phonon thermal conductance, respectively.
Discreteness of energy levels~\cite{Zianni}, Coulomb
interactions~\cite{Kubala,Swirkowicz}, interference
effects~\cite{Liu22}, and so on lead to the fact that nanodevices
have more thermal efficiency than bulky samples.
\par In recent years, investigation of the thermoelectrical
devices fabricated from quantum dots (QDs) has attracted a lot of
attention theoretically and
experimentally~\cite{Dubi,Andreev,Liu22,Dubi2,Xu,Ying,Ahmadian,Costi,Swirkowicz,Galperin,Tan}.
However, thermoelectric effects through double quantum dots
(DQDs) results in novel phenomena needing more studies. Few
articles have analyzed thermopower through a DQD in just recent
years~\cite{Zhang,Chi1,Trocha}. F. Chi and co-workers~\cite{Chi1}
showed $ZT$ has two huge peaks in the vicinity of the
electron-hole symmetry points. Trocha and J.
Barna\'{s}~\cite{Trocha} reported $ZT$ in DQD systems is enhanced
due to Coulomb interactions and interference effects. We have
recently studied~\cite{Bagheri} the effects of Coulomb
interactions and interdot tunneling on the thermopower of a DQD
system weakly coupled to metallic electrodes.
\par In this article, the influences of polarization of the leads,
temperature, and interdot tunneling on the figure of merit and
the thermoelectric conductances of a serial DQD weakly coupled to
ferromagnetic leads are analyzed. Many-body representation
introduced in Ref.~\cite{Fransson} is used to obtain formal
expressions for electrical and thermal conductances, thermopower,
and figure of merit. Hamiltonian and model used for calculations
are described in next section. Section 3 is devoted to numerical
results and in the end, some sentences are given as conclusion.

\section{Model and Method}  %%%  2.
We consider two single-level quantum dots coupled to ferromagnetic
leads. Hamiltonian describing the system is given as
\begin{align}\label{Eq.1}
H=&\sum_{\alpha k\sigma}\varepsilon_{\alpha
k\sigma}c^{\dag}_{\alpha k\sigma}c_{\alpha
k\sigma}+\sum_{i\sigma}\varepsilon_{i}n_{i\sigma}+\sum_{i}U_{i}n_{i\uparrow}n_{i\downarrow}+
U_{12}\sum_{\sigma\sigma'}n_{L\sigma}n_{R\sigma'}\\
\nonumber
&+t\sum_{\sigma}[d^{\dag}_{L\sigma}d_{R\sigma}+H.C]+\sum_{\alpha
k\sigma}[V_{\alpha k\sigma}c^{\dag}_{\alpha
k\sigma}d_{\alpha\sigma}+H.C]
\end{align}
where $c_{\alpha k\sigma}(c^{\dag}_{\alpha k\sigma})$ destroys
(creates) an electron with spin $\sigma$, wave vector $k$, in
lead $\alpha=L,R$. Energy levels of the leads are spin-dependent
because the leads are ferromagnetic. The second and third terms
in the above equation describe each dot.
$d_{i\sigma}(d^{\dag}_{i\sigma})$is annihilation (creation)
operator of the $i'$th ($i=L,R$) dot and
$n_{i\sigma}=d^{\dag}_{i\sigma}d_{i\sigma}$. Energy levels of each
dot are assumed to be degenerate. $U_{i}$ and $U_{12}$ denote,
respectively, on-site and interdot Coulomb repulsions. $t$ is the
interdot tunneling strength, whereas $V_{\alpha k\sigma}$ denotes
the dot-lead coupling strength. In order to study  the system,
the Hubbard operators have been used. In this model, the
Hamiltonian of the isolated DQD system is diagonalized and the
creation and annihilation operators are expanded in terms of
obtained eigenstates. It is obvious that the system has 16
different states shown as $|N,n>$ where $N=0...4$ is the number of
electrons, whereas $n$ denotes $n$'th state of $N$-electron
configuration. By expanding annihilation operator as
\begin{equation}\label{Eq.2}
  d_{\alpha\sigma}=\sum_{Nnn'}(d_{\alpha\sigma})_{NN+1}^{nn'}X_{NN+1}^{nn'}
\end{equation}
where the Hubbard operator, $X_{NN+1}^{nn'}=|Nn><N+1n'|$,
describes a transition in which an electron inside the DQD system
is annihilated, the whole Hamiltonian is given as
\begin{equation}\label{Eq.3}
  H=\sum_{\alpha k\sigma}\varepsilon_{\alpha k\sigma}c^{\dag}_{\alpha
  k\sigma}c_{\alpha k\sigma}+\sum_{Nn}E_{Nn}h_{N}^{n}+\sum_{\alpha k\sigma
  Nnn'}[V_{\alpha k\sigma}(d_{\alpha\sigma})_{NN+1}^{nn'}c^{\dag}_{\alpha k\sigma}X_{NN+1}^{nn'}+H.C]
\end{equation}
where $h_{N}^{n}=|Nn><Nn|$.
\par Now, population number $P_{Nn}$, the probability of being in
the state $|Nn>$, is computed by means of density matrix
approach. Coupling to the leads is considered so weak that
non-diagonal elements of the density matrix are ignored because
they are proportional to $V_{\alpha k\sigma}^4$. Using Markov
approximation and wide band limit, the time evaluation of the
population numbers is given as~\cite{Fransson}
\begin{subequations}\label{Eq.4}
\begin{align}
\frac{dP_{01}}{dt}&=\sum_{\alpha
n}[-\Gamma^{\alpha}_{|01>\rightarrow|1n>}P_{01}+\Gamma^{\alpha}_{|1n>\rightarrow|01>}P_{1n}]\\
\frac{dP_{Nn}}{dt}&=\sum_{\alpha
n'}-[\Gamma^{\alpha}_{|Nn>\rightarrow|N-1n'>}+\Gamma^{\alpha}_{|Nn>\rightarrow|N+1n'>}]P_{Nn}+\\
\nonumber & \Gamma^{\alpha}_{|N-1n'>\rightarrow|Nn>}P_{N-1n'}
+\Gamma^{\alpha}_{|N+1n'>\rightarrow|Nn>}P_{N+1n'}\\
\frac{dP_{41}}{dt}&=\sum_{\alpha
n}[-\Gamma^{\alpha}_{|41>\rightarrow|3n>}P_{41}+\Gamma^{\alpha}_{|3n>\rightarrow|41>}P_{3n}]
\end{align}
\end{subequations}
It is obvious that there is one configuration for zero- and
four-electron state. The transition rates are
\begin{subequations}\label{Eq.5}
\begin{align}
\Gamma^{\alpha}_{|Nn>\rightarrow|N+1n'>}&=\frac{1}{\hbar}\sum_{\sigma}\Gamma^{\alpha}_{\sigma}|(d_{\alpha\sigma})_{NN+1}^{nn'}|^2f_{\alpha}(E_{N+1n'}-E_{Nn})\\
\Gamma^{\alpha}_{|N+1n'>\rightarrow|Nn>}&=\frac{1}{\hbar}\sum_{\sigma}\Gamma^{\alpha}_{\sigma}|(d_{\alpha\sigma})_{NN+1}^{nn'}|^2f_{\alpha}^{-}(E_{N+1n'}-E_{Nn})
\end{align}
\end{subequations}
where $\Gamma^{\alpha}_{\sigma}=2\pi\sum_{k\in \alpha}|V_{\alpha
k\sigma }|^2$ is the spin-dependent tunneling rate,
$f_{\alpha}(x)=(1+exp((x-\mu_{\alpha})/kT_{\alpha}))^{-1}$ is the
Fermi-Dirac distribution function in which $\mu_\alpha$ and
$T_\alpha$ stand for the chemical potential and the operating
temperature of the lead $\alpha$, respectively, and
$f^{-}_{\alpha}=1-f_{\alpha}$. The charge and energy currents are
computed by solving the Eqs. \eqref{Eq.4} in the steady state
situation ($\frac{dP_{Nn}}{dt}=0$)
\begin{subequations}\label{Eq.6}
\begin{align}
  I^{\alpha}&=-e\sum_{N,N',n,n'}\Gamma^{\alpha}_{|Nn>\rightarrow|N'n'>}P_{|Nn>}sgn(N'-N)\\
  Q^{\alpha}&=\sum_{N,N',n,n'}\Gamma^{\alpha}_{|Nn>\rightarrow|N'n'>}(E_{|N'n'>}-E_{|Nn>}) P_{|Nn>}sgn(N'-N)
\end{align}
\end{subequations}
where $sgn(x)$ is a signum function.
\par To compute the thermoelectrical characteristics of the
system, the linear response regime is used. We assume that the
left lead is slightly hotter than the right one ($T_L=T_R+\Delta T
$), and $\mu_L=\mu_R-e\Delta V$, so that the charge and energy
currents are given as follows
\begin{subequations}\label{Eq.7}
\begin{align}
     I^{\alpha}&=G_{V}\Delta V+G_{T}\Delta{T}\\
    Q^{\alpha}&=M\Delta V+K\Delta{T}
\end{align}
\end{subequations}
where $G_V$ is the electrical conductance and $G_T$ is the
thermal coefficient. Thermopower is defined as minus ratio of
induced voltage to applied temperature gradient under condition
that the charge current is zero, so we have $S=\frac{G_T}{G_V}$.
Putting $I^{L}=1/2(I^L-I^R)$ and expanding the fermi-Dirac
distribution function as $f_{L}(x)=f_R(x)-x/Tf'(x)\Delta T+e\Delta
Vf'(x)$ where $f'(x)=\partial f(x)/\partial x$, one can easily
obtain:
\begin{subequations}\label{Eq.8}
\begin{align}
     G_V&=-\frac{e^2}{2\hbar}\sum_{Nnn'}\sum_{\sigma}P_{Nn}\Gamma^{L}_{\sigma}[|(d_{L\sigma})_{NN+1}^{nn'}|^2f'(E_{N+1n'}-E_{Nn})+\\\nonumber
     &\qquad \qquad|(d_{L\sigma})_{N-1N}^{n'n}|^2f'(E_{Nn}-E_{N-1n'})]\\
    G_T&=\frac{e}{2\hbar}\sum_{Nnn'}\sum_{\sigma}P_{Nn}\Gamma^{L}_{\sigma}[|(d_{L\sigma})_{NN+1}^{nn'}|^2\frac{(E_{N+1n'}-E_{Nn})}{T}f'(E_{N+1n'}-E_{Nn})+\\\nonumber
     &\qquad \qquad|(d_{L\sigma})_{N-1N}^{n'n}|^2\frac{(E_{Nn}-E_{N-1n'})}{T}f'(E_{Nn}-E_{N-1n'})]
\end{align}
\end{subequations}
Using above equations and Onsager relation, the thermal
conductance is computed as~\cite{Zianni}
\begin{equation}\label{Eq.9}
  \kappa_c=[K-S^2G_VT]
\end{equation}
where
\begin{align}\label{Eq.10}
  K&=-\frac{1}{2\hbar}\sum_{Nnn'}\sum_{\sigma}P_{Nn}\Gamma^{L}_{\sigma}[|(d_{L\sigma})_{NN+1}^{nn'}|^2\frac{(E_{N+1n'}-E_{Nn})^2}{T}f'(E_{N+1n'}-E_{Nn})+\\\nonumber
     &\qquad \qquad|(d_{L\sigma})_{N-1N}^{n'n}|^2\frac{(E_{Nn}-E_{N-1n'})^2}{T}f'(E_{Nn}-E_{N-1n'})]
\end{align}
\par For simulation purpose, we assume that
$\Gamma^L_\uparrow=\Gamma^R_\uparrow=\Gamma_0$ and spin-down
tunneling rate is equal to
$\Gamma^L_\downarrow=\Gamma^R_\downarrow=\alpha\Gamma_0$ where
$0\leq\alpha\leq 1$. It is obvious that $\alpha=1$ denotes normal
metallic electrodes whereas $\alpha=0$ stands for half-metal
leads. we also set $\kappa_{ph}=3\kappa_0$ where
$\kappa_0=\frac{\pi^2k_B^2}{3h}$T is the quantum of thermal
conductance~\cite{Rego} and, assume that the single electron
levels in the QDs are degenerate. In recent years, the phonon
contribution in the transport through QDs has been extensively
studied experimentally and
theoretically~\cite{LeRoy,Huttel,Galperin22}.

\section{Results and discussions}
Figure of merit ($ZT$) as a function of the QDs' energy level and
the temperature is plotted in fig. 1 for different $\alpha$s.
Similar plot was previously presented in Ref.~\cite{Swirkowicz}
about a QD coupled to Ferromagnetic electrodes or in
Ref.\cite{Trocha} about a DQD coupled to external electrodes.  One
can observe that increasing $\alpha$ results in the reduction of
$ZT$. On the other hand, $ZT$ has some peaks whose intensities
decrease with increase of temperature, but become wider. The
results also show that $ZT$ approaches zero in
$-3<\varepsilon_i<0$ and high temperature ($T>4k$). With respect
to the fact that $ZT$ is a function of the Seebeck coefficient,
the electrical conductance, and the thermal conductance, the
evaluation of them will help us realize the behavior of $ZT$.

\par Figure. 2a describes the electrical conductance as a function
of the energy level in different situations. From eq. (8a) it is
obvious that $G_V$ has a peak whenever the necessary energy for
transition from $N$-electron state to $N\pm 1$-electron state
equals to the Fermi energy of the leads ($\mu_\alpha=0$). In low
temperature ($T=2k$), $f'(x)$ is a Lorentzian function with
narrow width, thus the peaks and the Coulomb valleys are clearly
observed. There are four peaks because four electron can exist in
the system. With increase of the temperature, the intensity of
the peaks reduces whereas they become wider. Furthermore, the
Coulomb valleys are vanished because increase of the temperature
weakens the Coulomb blockade effect. It is straightforward to
show that $G_V=\frac{1+\alpha}{2}G_0$ where $G_0$ is the
electrical conductance of a double quantum dot coupled to  normal
metal electrodes. Therefore, the polarized electrodes decrease the
electrical conductance of the system. The thermal coefficient as
a function of energy is shown in fig. 2b. One can easily observe
from eq.(8b), for energies in which $G_V$ has a maximum $G_T$ and
as result, $S$ become zero because of $E_{Nn}\approx E_{N\pm
1n'}$. In these energy points, although the transition from
$N$-electron state to $N\pm 1$-electron state results in the
electrical current, no net energy transports in the process.
Indeed, the temperature gradient does not have any significant
role in producing the charge current in resonance energies. In
addition, the thermal coefficient becomes zero in some other
energies. From eq.(8b), it is obvious when the system is in a
certain state ($P_{Nn}\simeq 1$) and
$E_{N+1n'}-E_{Nn}=-(E_{Nn}-E_{N-1n'})$, then $G_T=0$. These
energy points are so-called electron-hole symmetry points and the
effect has been recently studied in a multilevel QD~\cite{Liu1}.
In symmetry points, the current transported from the lead to the
QD is equal to the current transported in opposite direction, so
that the thermopower becomes zero. Indeed, both electrons and
holes participate in producing the current with different signs.
It is worth noting that $G_T$ goes from positive to negative
values in resonance points whereas it behaves reversely in
symmetry points. This effect, bipolar effect, has been recently
reported in a multilevel quantum dot~\cite{Liu1}. It comes from
the fact that in each side of it, a kind of charge carrier
(electron or hole) participates in transport. This leads to the
oscillation of the thermopower. Increasing temperature gives rise
to widening $G_T$ but its intensity will be reduced. The
influence of the polarization of the leads on the thermal
coefficient is the same as $G_V$ and as a result, the thermopower
is independent of the $\alpha$. This result was previously
reported about a single level QD coupled to ferromagnetic
leads~\cite{Dubi2}. However, results presented in
Ref.~\cite{Swirkowicz} show that the thermopower weakly depends
on the polarization of the leads. This difference comes from the
different models used for studying the system. Indeed, in
nonequilibrium Green's function formalism used in
Ref.~\cite{Swirkowicz}, the broadening of the QD level due to
coupling between the QD and the leads is taken into account. That
leads to the dependence of the thermopower on the magnetic
polarization. In rate equations approach, the broadening of the
QD level due to coupling is disregarded. Figure of merit is zero
in resonance and symmetry pints as one can see in fig. 1 due to
$ZT\propto S$. Furthermore, $ZT$ decreases with increase of
temperature because increasing temperature results in reducing
$S$ and $G_V$. On the other hand, although the thermopower is
spin-independent, the figure of merit is reduced by increase of
polarization because of $ZT\propto G_V$.

\par The thermal  conductance as a function of temperature and
energy level is shown in fig. 3. Results show that the thermal
conductance takes the highest values in electron-hole symmetry
points when the temperature is sufficiently high. In these points,
electrons and holes participate in the transfer of charge and heat
with the same weight. Because the current carried by holes is in
the opposite direction with one carried by electrons, the charge
current becomes zero in these energies so that one can observe as
valleys of $G_V$ in fig. 2a. In contrast with the charge current,
electrons and holes carry the energy in the same direction
resulting in the appearance of the peaks in $\kappa_c$. As
expect, increasing temperature gives rise to the increase of
$\kappa_c$.  In low temperature, the peaks of $G_V$ and
$\kappa_c$ are happened in nearly same energies. Indeed, the
significant difference between $G_V$ and $\kappa_c$ is observed
in high temperature. In low temperature, when the transition
energy is equal to resonant energy, one peak develops in both
$G_V$ and $\kappa_c$, resulting due to the transfer of current
and heat through the system by electrons~\cite{Swirkowicz,Trocha}.

\par Figure. 4 describes figure of merit as a function of the
polarization of the leads and the interdot tunneling strength. It
is observed that increase of the polarization of the leads
results in the reduction of $ZT$. Results also show that the
interdot tunneling leads to decrease of $ZT$. Indeed, From fig. 4
one can approximately consider $ZT\propto \alpha t$ so that $ZT$
obtained from strong $\alpha$ (weak polarization) and weak $t$ is
equal to result obtained from weak $\alpha$ (strong polarization)
and strong $t$. It is important to note that the eigenvalues of
the DQD are a function of the interdot tunneling, so the position
of symmetry and resonance points depends on the interdot tunneling
 strength. It is worth noting that increase of interdot tunneling results in reduction of $G_V$~\cite{Bagheri}.
  Furthermore, although the thermopower is
spin-independent, the electrical conductance is reduced by
increasing polarization, and as a result the figure of merit
reduces when the polarization of leads increases. In addition,
intra- and interdot Coulomb repulsions affect the figure of merit
significantly. Recently, we have shown that the increase of
Coulomb repulsion results in the increase of figure of
merit~\cite{Bagheri}. Indeed, the level spacing is increased by
increase of Coulomb repulsion and as a result, the bipolar effect
will be reduced. The same effect was previously reported about a
multilevel QD~\cite{Liu1}.

\section{Conclusion}
\label{conclusion} In this paper, the thermopower in a serial
double quantum dot system weakly coupled to ferromagnetic leads is
studied using Hubbard operators. Formal expressions for thermal
conductance, electrical conductance, and thermal coefficient are
obtained using density matrix approach. The effect of
temperature, interdot tunneling, and polarization of leads on the
thermoelectrical characteristics of the system is examined.
Results show that increase of temperature results in decrease of
electrical conductance intensity whereas it becomes wider. It is
found that although the thermopower is independent of
polarization, the figure of merit is reduced by increase of
polarization. Influence of the interdot tunneling on the figure
of merit is also analyzed.

\bibliographystyle{model1a-num-names}
\bibliography{<your-bib-database>}

\begin{thebibliography}{00}

\bibitem{Dubi}
Y. Dubi and M. Di Ventra, Rev. Mod. Phys 83 (2011) 131.


\bibitem{Andreev}
A.V. Andreev, K. A. Matveev, Phys. Rev. Lett. 86 (2001) 280.

\bibitem{Vashaee}
D. Vashaee and A. Shakouri, Phys. Rev. Lett, 92 (2004) 106103.

\bibitem{Finch}
C. M. Finch, V. M. Garc\'{i}a-Su\'{a}rez,  C. J. Lambert, Phys.
Rev. B 79 (2009) 033405.

\bibitem{Dubi1}
Y. Dubi and M. Di Ventra, Phys. Rev. B 79 (2009) 115415.

\bibitem{Soulier}
C. Bera, M. Soulier, C. Navone, G. Roux, J. Simon, S. Volz, N.
Mingo, J. Appl. Phys. 108 (2010) 124306.

\bibitem{Nozaki}
D. Nozaki, H. Sevinçli, W. Li, R. Gutiérrez, G. Cuniberti, Phys.
Rev. B 81 (2010) 235406.

\bibitem{Kuo}
D. M.-T. Kuo, and Y-C. Chang, Phys. Rev. B 81 (2010) 205321.

\bibitem{Moca}
C. P. Moca, A. Roman, D. C. Marinescu, Phys. Rev. B 83 (2011)
245308.

\bibitem{Singh}
D. Singh, J. Y. Murthy,  T. S. Fisher, J. Apll. Phys 110 (2011)
044317.


\bibitem{Hochbaum}
A. I. Hochbaum, R. Chen, R. D. Delgado, W. Liang, E. C. Garnett,
M. Najarian, A. Majumdar, P. Yang, Nature, 451 (2008) 163.

\bibitem{Balandin}
A. A. Balandin,  O. L. Lazarenkova, Appl. Phys. Lett. 82 (2003)
415.
\bibitem{Khitun}
A. Khitun, A. Balandin, K.L. Wang, G. Chen, Physica E, 8 (2000)
13.

\bibitem{Mazzamuto}
F. Mazzamuto, V. Hung Nguyen, Y. Apertet, C. Ca\"{e}r, C.
Chassat, J. Saint-Martin, P. Dollfus, Phys. Rev. B 83 (2011)
235426.

\bibitem{Kubala}
B. Kubala, J. K\"{o}nig, J. Pekola, Phys. Rev. Lett. 100 (2008)
066801.

\bibitem{Garg}
A. Garg, D. Rasch, E. Shimshoni,  A. Rosch, Phys. Rev. Lett. 103
(2009) 096402.

\bibitem{Zianni}
X. Zianni, Phys. Rev. B 78 (2008) 165327.

\bibitem{Swirkowicz}
R. \'{S}wirkowicz, M. Wierzbicki, J. Barna\'{s}, Phys. Rev. B 80
(2009) 195409.

\bibitem{Liu22}
Y. S. Liu, D. B. Zhang, X. F. Yang, J. F. Feng, Nanotechnology 22
(2011) 225201.

\bibitem{Dubi2}
Y Dubi, M. Di Ventra, Phys. Rev. B 79 (2009) 081302R.

\bibitem{Xu}
Y. Xu, X. Chen, J-S. Wang, B-L. Gu,  W. Duan, Phys. Rev. B 81
(2010) 195425.

\bibitem{Ying}
Y, Ying, G. Jin, Appl. Phys. Lett. 96 (2010) 093104.

\bibitem{Ahmadian}
Y. Ahmadian, G. Catelani,  I. L. Aleiner, Phys. Rev. B 72 (2005)
245315.

\bibitem{Costi}
T. A. Costi, V. Zlati\'{c}, Phys. Rev. B 81 (2010) 235127.



\bibitem{Galperin}
M. Galperin, A. Nitzan, M. A. Ratner,  Phys. Rev. B 75 (2007)
155312.

\bibitem{Tan}
A. Tan, S. Sadat, P. Reddy,  Appl. Phys. Lett. 96 (2010) 013110.


\bibitem{Zhang}
Z-Y. Zhang, J. Phys: Condensed matter, 19 (2007) 086214.

\bibitem{Chi1}
F. Chi, J. Zheng, X. D. Lu, K. C. Zhang, Phys. Lett. A, 375
(2011) 1352.

\bibitem{Trocha}
P. Trocha,  J. Barna\'{s}, arXive:1108.2422v1, (2011).

\bibitem{Bagheri}
M. B. Tagani, H. R. Soleimani, Physica B 407 (2012) 765.

\bibitem{Fransson}
J. Fransson, M.R\.{°a}sander, Phys. Rev. B, 73 (2006) 205333.



\bibitem{Rego}
L. G. C. Rego, G. Kirczenow, Phys. Rev. Lett. 81 (1998) 232.

\bibitem{LeRoy}
 B.J. LeRoy, S.G. Lemay, J. Kong, and C. Dekker,
Nature, 432 (2004) 371.

\bibitem{Huttel}
A. K. H\"{u}ttel, B. Witkamp, M. Leijnse, M. R. Wegewijs,  H. S.
J. van der Zant, Phys. Rev. Lett. 102 (2009) 225501.

\bibitem{Galperin22}
M, Galperin, M. A. Ratner, A. Nitzan, J. Phys.: Condens. Matter
19 (2007) 103201.

\bibitem{Liu1}
J. Liu, Q. F. Sun,  X. C. Xie, Phys. Rev. B, 81 (2010) 245323.

 \end{thebibliography}

\newpage
\textbf{Figure captions}

Figure. 1: Color map of figure of merit. Parameters are: $t=0.2
meV$, $U_i=2 meV$,  $U_{12}=1 meV$ and $\Gamma_0=10\mu{eV}$.

\par
Figure. 2: $G_V$ and $G_T$ as a function of energy. Solid
($T=2K,\alpha=1$), dashed ($T=5K,\alpha=1$), dash-dotted
($T=2K,\alpha=0.5$), and dotted ($T=5K,\alpha=0.5$). Other
parameters are the same as fig. 1 .

\par
Figure. 3: Color map of thermal conductance. Parameters are the
same as fig. 1.

\par
Figure. 4: Figure of merit as a function of polarization and
interdot tunneling. Parameters are the same as fig. 1 except
$T=2K$ and $\varepsilon_i=-1.5meV$.

\newpage

\begin{figure*}
\includegraphics[width=1\textwidth]{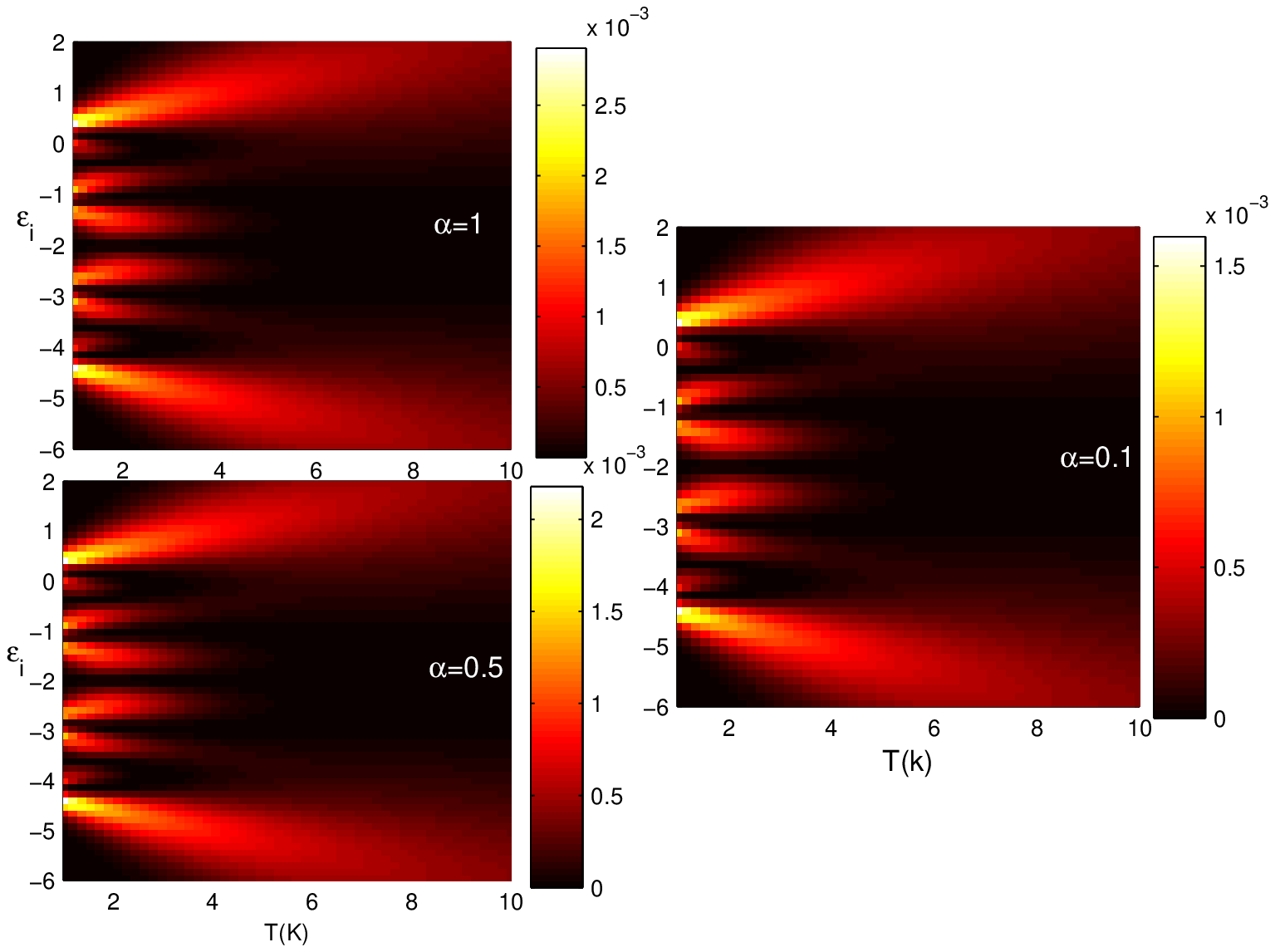}\nonumber
$Figure. 1$
\end{figure*}

\begin{figure*}
\includegraphics[width=1\textwidth]{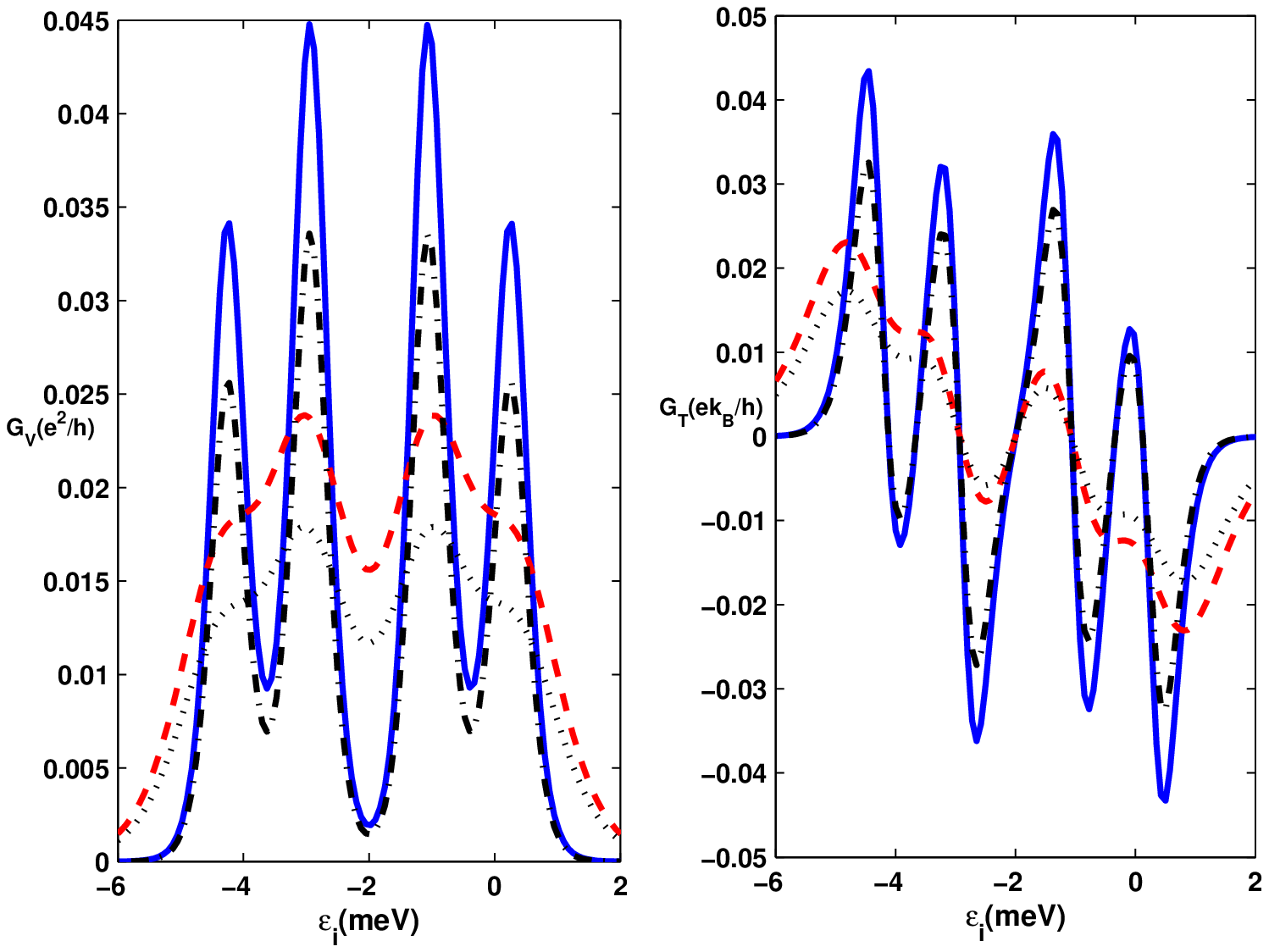}
$Figure. 2$
\end{figure*}

\begin{figure*}
\includegraphics[width=1\textwidth]{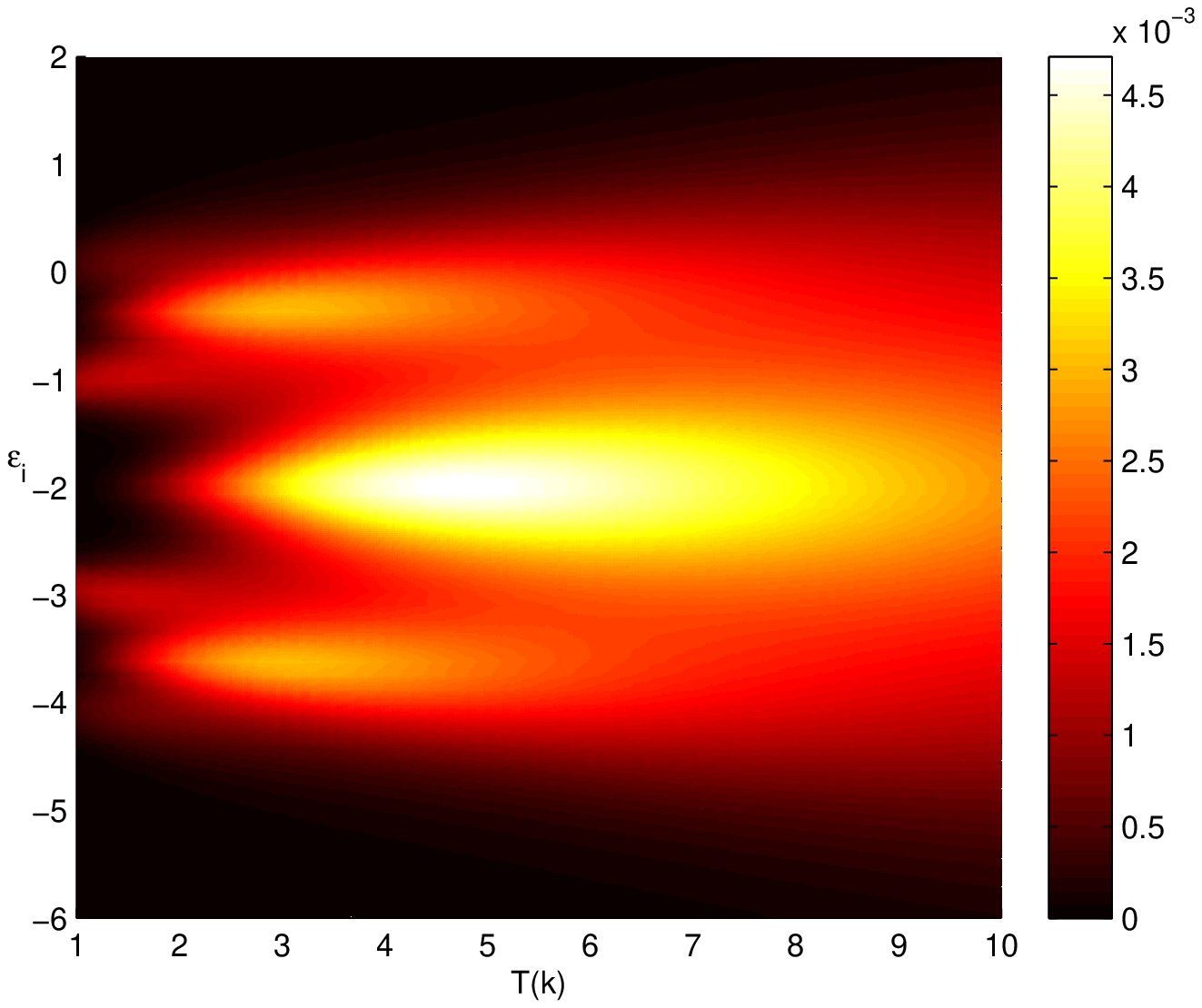}
$Figure. 3$
\end{figure*}

\begin{figure*}
\includegraphics[width=1\textwidth]{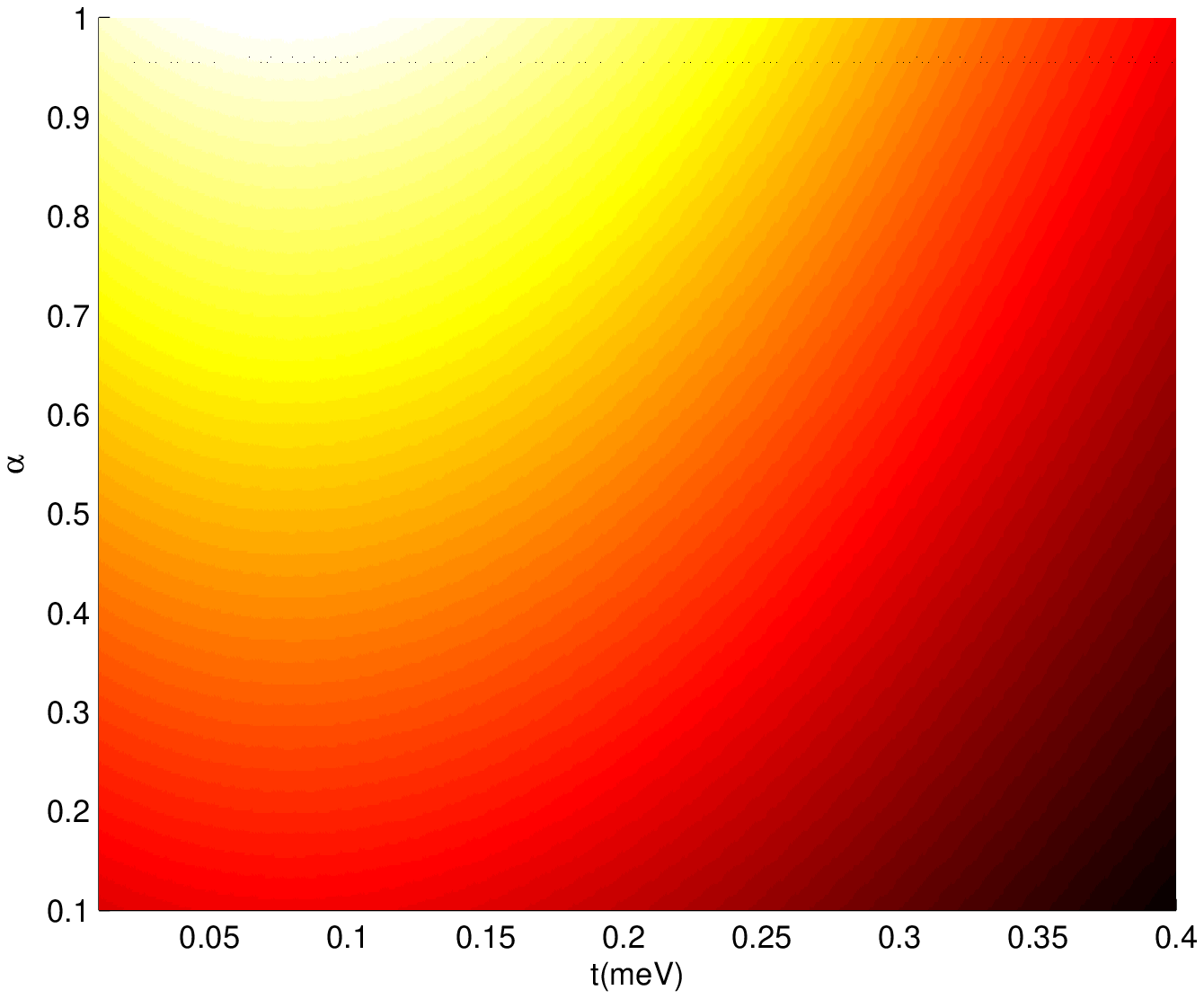}
$Figure. 4$
\end{figure*}

\end{document}